# Prediction of Macrobending and Splice Losses for Photonic Crystal Fibers based on the Effective Index Method

G. S. KLIROS, J. KONSTANTINIDIS, C. THRASKIAS
Department of Aeronautical Sciences
Div. of Electronics & Communication Engineering
Hellenic Air-force Academy
GREECE

*Abstract:* - An index-guiding photonic crystal fiber (PCF) with an array of air holes surrounding the silica core region has special characteristics compared to conventional single-mode fibers (SMFs). Using the effective index method and the Gaussian beam propagation theory, the macro-bending and splice losses for PCFs are investigated. The wavelength dependence of the cladding index of the PCF has been taken properly into account. We obtain the effective spot size for different configurations of PCFs, which is used for computing the splice losses. The Gaussian approximation for the fundamental modal field leads to simple closed-form expressions for the splice losses produced by transverse, longitudinal and angular offsets. Calculations of macro-bending losses are based on antenna theory for bend standard fibers.

## 1 Introduction

Photonic crystal fiber (PCF) is a unique type of optical fiber [1]-[3] incorporating an array of air holes that run along its length, reminiscent of a crystal lattice, which gives to this type of fiber, its name. There are two main types of PCF: air-guiding which guides light via a photonic band-gap effect and index-guiding which guides via a modified total internal reflection mechanism.

In air-guided PCF, the core is hollow, and light is guided by the photonic band gap (PBG) effect, a mechanism that does not require a higher refractive index in the core in order to confine and guide light. The PBG guidance effect relies on coherent backscattering of light into the core.

In index-guided PCFs the core area is solid and the light is confined to a central core as in conventional fibers. A PCF consists of a pure silica fiber with an array of air holes running the full length of the fiber. The core is formed by omitting the central air hole of the structure (Fig.1). The lower effective refractive index of the surrounding holes forms the cladding resulting in an index guidance mechanism analogous to total internal reflection.

PCFs have been shown to possess many significant properties like single mode operation over wide range of wavelengths, highly tunable dispersion, propagation of high power densities without exciting unwanted non-linear effects and high birefringence. Such properties are of practical importance in the design of sophisticated broadband optical telecommunication networks [4] and active sensor systems [5].

The arrangement and spacing of air holes in a PCF provide freedom to tailor the dispersion as well as the coupling properties for telecommunication and sensing applications [6]. As the performance of PCFs improves and new integrated designs using the existing optical fiber technology are demonstrated, the study of splicing characteristics of a PCF to conventional single mode fibers or to a PCF of different hole spacing, is very important [7],[8].

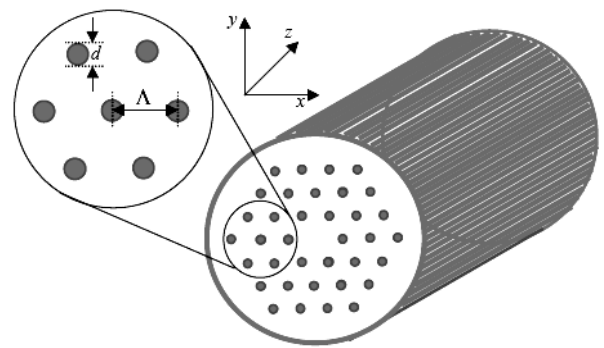

Fig.1: Schematic representation of a PCF showing the hole diameter d and hole to hole spacing (pitch) $\Lambda$.

In this paper, we present analytical calculations of macro-bending and splice losses for index-guiding PCFs. We consider the splice loss due to lateral, longitudinal as well as angular offsets, when a photonic crystal fiber (PCF) is spliced to an identical PCF or to a step index single mode fiber (SMF).

Our treatment is based on the effective index method and the Gaussian beam propagation theory and provides closed form expressions for macrobending and splice losses for PCFs taking into account the material dispersion of silica. The case of splicing between a conventional SMF and a PCF is also considered and a strong dependence on the air hole spacing and wavelength is observed.

The paper is organized as follows: In section 2 the effective index model (EIM) is briefly presented and is used in order to calculate the cladding effective index. In section 3, the spot size (Mode Field Radius of the fundamental mode) is calculated analytically. Predictions of macrobending and splice losses based on easy-to-evaluate closed-form expressions are presented in sections 4 and 5, respectively. We conclude with a summarizing Section 6.

## 2 The effective index model

In the analysis of a PCF, the effective cladding index is an important parameter; it facilitates our understanding of PCFs and allows the application of conventional theories of fiber propagation without necessarily resorting to detailed numerical models. The effective cladding index can be obtained by investigating the propagation of the fundamental cladding mode. The fundamental cladding mode (space filling mode) is defined to be the mode that exhibits the highest effective index.

In the effective index method, the photonic crystal cladding is replaced by a single material having refractive index equal to the modal index of fundamental space filling mode (FSFM). The effective index of the FSFM, $n_{eff,cl} = \beta_{FSM}/k_0$, ($k_0 = 2\pi/\lambda$, is the free space propagation constant with wavelength $\lambda$), is calculated by solving the scalar wave equation within a unit cell of Photonic Crystal, centered on one of the holes. The diameter of these cells equals the pitch $\Lambda$, between the holes and their hexagonal shape has been approximated by a circular one in order to make a circular symmetric mode solution possible. The solution $\Psi(R)$ of the scalar wave equation in the air hole and silica regions is given by

$$\Psi(R) = AI_0(WR), \quad R < 1$$
$$= BJ_0(RU) + CY_0(RU) \quad R < 1 \quad (1)$$

where $R = r/a$, $a$ - being the radius of air hole. Using boundary conditions for the cladding mode-field and its normal derivative at the edges of each unit cell, the following eigenvalue equation is obtained:

$$\frac{I_1(W)}{(UW)^2}(W-U)\left[J_0(U)Y_1(U) - Y_0(U)J_1(U)\right]$$
$$\cdot \{I_0(W)\varepsilon_2 W\left[J_1(u)Y_1(U) - J_1(U)Y_1(u)\right] -$$
$$-\varepsilon_1 I_1(W)U\left[J_0(U)Y_1(u) + J_1(u)Y_0(U)\right]\} = 0 \quad (2)$$

where the parameters U, W and u are defined as follows

$$\left. \begin{array}{l} U = k_0 a\sqrt{n_s^2 - n_{cl,eff}^2} \\ W = k_0 a\sqrt{n_{cl,eff}^2 - n_{air}^2} \\ u = k_0 b\sqrt{n_s^2 - n_{cl,eff}^2} \end{array} \right\} \quad (3)$$

and $b = \Lambda\sqrt{\sqrt{3}/(2\pi)}$ is the radius of the outer circle, which is obtained by equating the filling fraction of hexagonal unit cell and its circular approximation, $n_s$, $n_{air}$ are the refractive indices of pure silica and air, respectively.

Having determined the effective index of the cladding $n_{eff,cl}$, we can calculate the propagation constant and hence, the modal effective index $n_{eff}$ of guided wave in PCF is obtained similar to step index fiber with core index $n_{co} = n_s$, core radius $a_{eff} = 0.64\Lambda$ [6], and cladding index $n_{cl} = n_{eff,cl}$. The wavelength dependence of refractive index for silica is taken into account through the Sellmeier's formula. The fundamental mode solution for the equivalent step index fiber is written as

$$\Psi(R) = \frac{1}{J_0(U_{eff})}J_1(RU_{eff}), \quad R < 1$$
$$= \frac{1}{K_0(W_{eff})}K_1(RW_{eff}), \quad R > 1 \quad (4)$$

where now the eigenvalue equation obtained is similar to the eigenvalue equation of step index fiber except the waveguide parameters which are

$$\left. \begin{array}{l} U_{eff} = k_0 a_{eff}\sqrt{n_s^2 - n_{eff}^2} \\ W_{eff} = k_0 a_{eff}\sqrt{n_{eff}^2 - n_{cl,eff}^2} \\ V_{eff} = k_0 a_{eff}\sqrt{n_s^2 - n_{cl,eff}^2} \end{array} \right\} \quad (5)$$

Fig. 2 shows the cladding effective index as a function of wavelength $\lambda$ for $d/\Lambda = 0.35$. In the wavelength range $\lambda \leq 1.7\mu m$ the numerical results can be reasonably fitted by the expression [9]:

$$n_{eff} \cong \bar{n} + (n_S - \bar{n})\cosh^{-2}(\lambda/\Lambda) \quad (6)$$

where
$$\bar{n} \equiv \lim_{\lambda \gg \Lambda} n_{cl,eff} = f n_{air} + (1-f)n_S \quad (7)$$

and $f = \pi/(2\sqrt{3})(d/\Lambda)^2$ is the filling fraction.

Mortensen et al. proposed another effective V-parameter for PCFs [10]. However, this definition $V_{eff}$ in Eq. 5 is intrinsically different from the original V parameter definition in Step-Index Fiber theory and corresponds to the normalized transverse attenuation constant W-parameter.

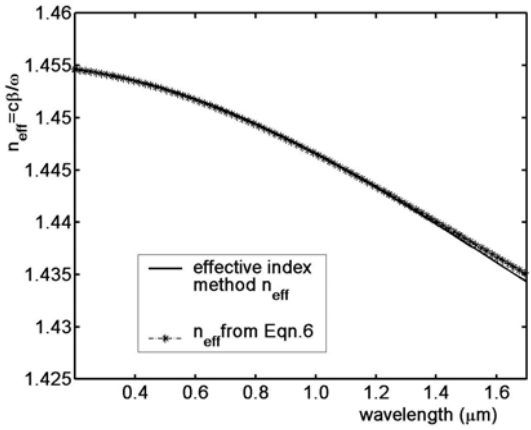

Fig. 2: Effective index of the fundamental mode versus wavelength λ for PCF with d/Λ=0.3.

## 3 Effective modal spot size

Spot size is another important parameter in the context of beam divergence, splice loss, bending loss and source to fiber coupling efficiency and nonlinearity coefficient. Petermann's definition [11] of spot size *w* can be applied to close-to-Gaussian modes, like the fundamental mode of the PCFs [12].

$$\frac{w}{a_{eff}} = \left[\frac{2\int_0^\infty R^3 \Psi^2(R) dR}{\int_0^\infty R \Psi^2(R) dR}\right]^{1/2} \quad (8)$$

Using Equation (4) in (8) the modal spot size w normalized to pitch Λ is calculated as

$$\frac{w}{\Lambda} = \sqrt{2}\frac{a_{eff}}{\Lambda}\left[\frac{I_1 + I_3}{I_2 + I_4}\right]^{1/2} \quad (9)$$

where we have explicitly evaluated the following integrals:

$$I_1 = \frac{1}{J_0^2(U_{eff})}\int_0^1 R^3 J_1^2(U_{eff}R)dR = \\ = \frac{J_1(U_{eff})^2 + J_2(U_{eff})^2}{6J_0^2(U_{eff})} \quad (10)$$

$$I_2 = \frac{1}{J_0^2(U_{eff})}\int_1^\infty R J_1^2(U_{eff}R)dR \\ = \frac{J_1(U_{eff})^2 - J_0(U_{eff})J_2(U_{eff})}{2J_0^2(U_{eff})} \quad (11)$$

$$I_3 = \frac{1}{K_0^2(W_{eff})}\int_1^\infty R^3 K_1^2(W_{eff}R)dR \\ = \frac{K_2(W_{eff})^2 - K_1(W_{eff})^2}{6K_0^2(W_{eff})} \quad (12)$$

and

$$I_4 = \frac{1}{K_0^2(W_{eff})}\int_1^\infty R K_1^2(W_{eff}R)dR \\ = \frac{K_0(W_{eff})K_2(W_{eff}) - K_1(W_{eff})^2}{2K_0^2(W_{eff})} \quad (13)$$

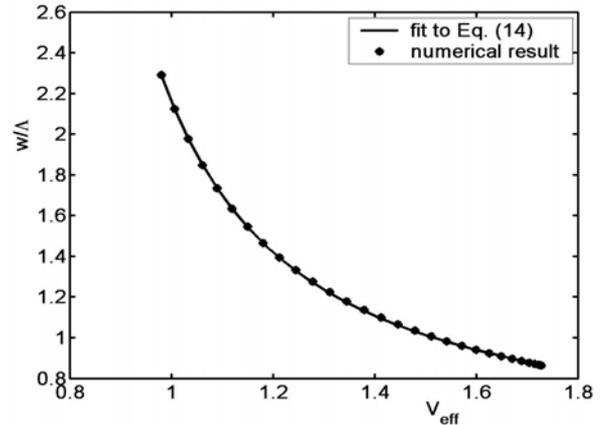

Fig. 3: Effective spot size of the fundamental mode versus effective normalized frequency for PCF with d/Λ=0.3 and hole separation Λ=2.3 μm.

Knowledge of effective spot size is an important starting point in the context of splice loss calculations. Due to the high index contrast between silica and air, the PCF can offer smaller spot sizes as compared to conventional fibers, which is useful for broad-band super-continuum generation, soliton pulses transmission and optical signal processing. On the other hand, PCF can have large-mode-area with low nonlinearity due to their endlessly single-mode properties combined with large spot sizes.

Fig. 3 shows the calculated modal spot size (dots) normalized to the air-hole pitch $\Lambda$ as a function of the effective normalized frequency $V_{eff}$, along with a numerical fit through these points (solid line). As fitting function we choose to use the same general expression as used for the standard fibers [12]

$$\frac{w}{\Lambda} = \frac{A}{V_{eff}^{2/(2+g)}} + \frac{B}{V_{eff}^{3/2}} + \frac{C}{V_{eff}^6} \quad (14)$$

where g can be any positive value. The values of the fitting parameters are determined to be A=0.489, B=0.913, C=0.762 and g = 8. Calculations for different ratios d/$\Lambda$, fall on the same curve indicating that the spot size for a PCF is in fact a function of the parameter $V_{eff}$ only.

As it is shown in Fig. 3, the modal spot size decreases as the effective normalized frequency increases or the wavelength $\lambda$ decreases. This is due to the fact that at longer wavelengths the field penetrates into the air holes, reducing the cladding index and hence increasing the refractive index contrast.

## 3 Macrobending Loss

The investigation of macro-bending losses of PCFs is very important, not only from a practical handling point of view, but also plays a central role when defining the spectral window in which the fiber may be operated. Predictions of macro-bending loss in PCFs have been made using the antenna theory of Sakai and Kimura [13] or a phenomenological model within the tilted-index representation [14], [15]. The tilted index model is an exact mapping for a scalar description which works well for large mode area PCFs with $\lambda<<\Lambda$.

Here, we apply the standard radiation model [16] for bend standard fibers, making a full transformation of conventional SIF parameters such as $\Delta$, W, and V to fiber parameters appropriate to index-guiding PCFs. Thus, the bending-loss formula for the power attenuation coefficient of conventional SIF due to macro-bending is transformed to the following expression:

$$a\left(\frac{dB}{m}\right) = 4.343 \left(\frac{\pi}{4a_{eff}R}\right)^{1/2} \left(\frac{1}{W_{eff}}\right)^{3/2} \cdot \exp\left(-\frac{4RW_{eff}^3 \Delta_{eff}}{3a_{eff}V_{eff}^2}\right) \left(\frac{U_{eff}}{V_{eff}K_1(W_{eff})}\right)^2 \quad (15)$$

where R denotes the radius of the curvature in the bend and $\Delta_{eff}$ is the relative difference between the core and the effective cladding indices. Bending loss for different structures of PCF and two different bend radius R=6 cm (dispersion-compensating fiber coil) and R=12 cm, is calculated and shown in Figures 4a and 4b, respectively.

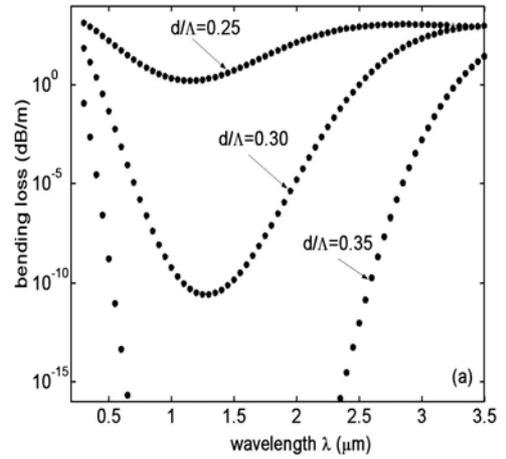

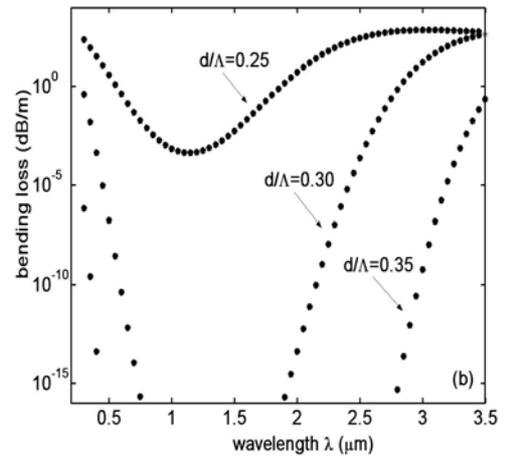

Fig. 4: Macro-bending loss for PCF with hole separation $\Lambda$=2.3 μm versus wavelength for different values of relative hole sizes d/$\Lambda$ μm and bending radius a) R=6 cm and b) R=12 cm.

We observe in accordance to previously published calculations [14] that the spectral width for minimum bend loss and the position of the operational window depends on the bending radius and the relative air hole size. A broad range of operating wavelengths for minimum bend loss is found for different configurations of PCFs. A larger hole-size results in broader window, whereas the pitch determines the center position of the window. As it is seen in Figs. 4a and 4b, the minimum bending loss appears approximately at a wavelength around $\Lambda/2$. More specifically, the figures reveal that for the realization of significantly bending-resistant PCFs, the normalized hole-size, d/$\Lambda$, should be around 0.30 or larger.

The critical bend radius is defined as the radius under which the fiber may not be bent in order for the excess macrobending loss to be below a given limit. We may derive the parametric dependence of the critical bend radius $R_C$ from the exponential factor of Eq. (15). The exponential factor increases dramatically when the argument is less than unity and thus we may define a critical bend radius from the following relation:

$$\frac{4RW_{eff}^3 \Delta_{eff}}{3a_{eff}V_{eff}^2} \simeq 1 \quad \text{or}$$

$$R_c \simeq \frac{4W_{eff}^3 \Delta_{eff}}{3a_{eff}V_{eff}^2} \propto \frac{k_0^2}{(\beta^2 - \beta_{FSM}^2)^{3/2}} \quad (16)$$

which can be written as

$$R_c \propto \frac{\Lambda^3/\lambda^2}{\Lambda^3(\beta^2 - \beta_{FSM}^2)^{3/2}} = \frac{1}{V_{PCF}^3}\frac{\Lambda^3}{\lambda^2} \quad (17)$$

where $V_{PCF} = \Lambda(\beta^2 - \beta_{FSM}^2)^{1/2}$ is the V-parameter introduced by Mortensen et al. [10], $\beta$ is the propagation constant of the fundamental mode and similarly $\beta_{FSM}$ is the propagation constant of fundamental space filling mode (FSFM). For index-guiding PCFs, the wavelength dependence of $\beta$ and $\beta_{FSM}$ gives rise to both a long-wavelength bend-edge (as in conventional fibers) as well as a short-wavelength bend-edge. In the limit of short wavelengths, the parameter $V_{PCF}$ tends to a constant [17] and therefore, $R_C \sim 1/\lambda^2$ which has been confirmed experimentally for a particular PCF design [18].

## 5 Calculation of Splice Losses

As we mentioned, a PCF has a two-dimensional cross-sectional structure in which the solid pure silica core is surrounded by a cladding region with air holes. If the fiber is left unsealed, the fiber may absorb unwanted liquids or gases. The best way to avoid this, is to splice the PCF to a standard fiber. In addition, splicing PCF to other PCF or standard SMF is important to many applications in optical telecommunications and integrated sensor systems. Therefore, the evaluation of splice losses that occur due to misalignment when two fibers are spliced, is an interesting issue.

In the following analysis, we present closed-form expressions for splice losses due to transverse, longitudinal and angular offsets using an overlapping-mode integration method and the Gaussian propagation theory. Such overlapping-mode integral calculations have been applied to the coupling analysis between SI-SMFs as well as thermally diffused expanded core (TEC) fibers [19]. When the air-filling factor of a PCF is smaller than 0.45, it is known that the modal field tends to have a Gaussian shape [20] and therefore, a Gaussian approximation for the modal fields is applicable.

The spot size of SI-SMFs is calculated using the Petermann's formula [12]:

$$w_P = w_M - (0.016 + 1.561V^{-7})\alpha \quad (18)$$

where $w_M$ is the Marcuse spot size given by

$$w_M = (0.650 + 1.619V^{-3/2} + 2.879V^{-6})\alpha \quad (19)$$

Equation (16) is accurate to 1% in the range of our interest 1.5<V<2.5.

The splice loss, when there are lateral offsets $d_x$ and $d_y$ in the x and y directions respectively, is expressed as:

$$\eta_{offset} = \frac{4w_1^2 w_2^2}{(w_1^2 + w_2^2)^2} e^{-\frac{2r_d^2}{w_1^2 + w_2^2}} \quad (20)$$

where $r_d = (d_x^2 + d_y^2)^{1/2}$ is the transverse offset between two fibers and $w_1$, $w_2$ are their spot sizes. Fig. 5 shows the splice loss between two identical PCFs in (dB) versus transverse offset for different values of ratio d/$\Lambda$ and for operating length $\lambda$=1.55 μm. As the ratio d/$\Lambda$ increases, the splice loss decreases drastically. Thus, the tolerance of transverse offset is less than 1 μm if the allowed loss is 0.5 dB. The above results are in good agreement with the predictions from numerical analysis [21]. It is clear from Fig. 5 that the loss dominates as we go to large air hole size in cladding.

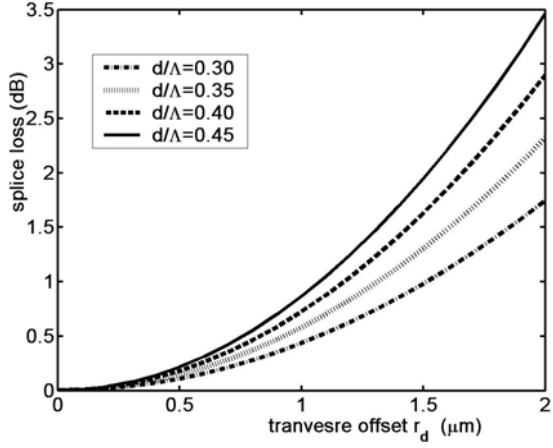

Fig. 5: Splice loss between PCF-fibers in (dB) versus transverse offset of ratio d/Λ at λ=1.55 μm. the hole separation is Λ=2.3 μm.

The influence of pitch on the splice loss due to transverse offset is also observed and shown in the Fig. 6. We see that the splice is large when the spacing between the holes are less and it decreases as spacing between air holes increases. It is due to the fact that when the air holes are at a large separation, the field is confined to core region only and it avoids the air-holes.

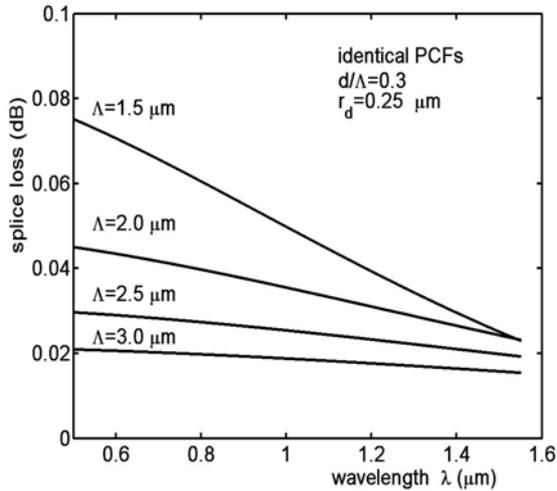

Fig. 6: Splice loss between identical PCF-fibers versus wavelength for different hole pitches Λ. The transverse offset was taken $r_d = 0.25$ μm.

The investigation of splice losses produced by a longitudinal offset between two PCF, is relevant to the design of micro-optical devices as lens-less in-line filters, integration of fiber-optic lenses and micro-mechanical fiber-optic switches.

When there is a longitudinal offset $z$ between the fibers, the splice loss can be expressed as:

$$\eta_z = \frac{4 w_1^2 w_2^2}{\frac{z^2}{k_0^2} + (w_1^2 + w_2^2)^2} \qquad (21)$$

Fig.7 shows the splice loss in (dB) versus longitudinal offset $z$ for the same parameters as in Fig. 5. As the ratio $d/\Lambda$ increases, the splice loss decreases drastically and as long as the offset is less than 4 μm, the loss is less than 0.01 dB. This result important for free-space interconnections in optical switching systems.

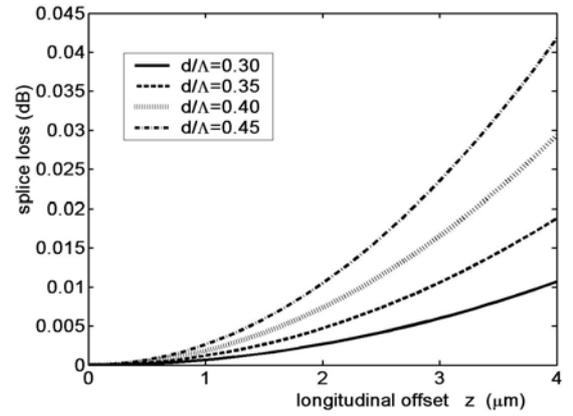

Fig.7: Splice loss between two identical PCF versus longitudinal offset for the parameters as in Fig. 5.

Splice losses due to angular offset is significant for designing high performance optoelectronic components and to fiber-optic angular alignment automation. When there is an angular misalignment $\theta$ between two fibers, the splice loss can be expressed as follows:

$$\eta_\theta = \frac{4 w_1^2 w_2^2}{(w_1^2 + w_2^2)^2} e^{-\frac{k^2 \theta^2 w_1^2 w_2^2}{2(w_1^2 + w_2^2)}} \qquad (22)$$

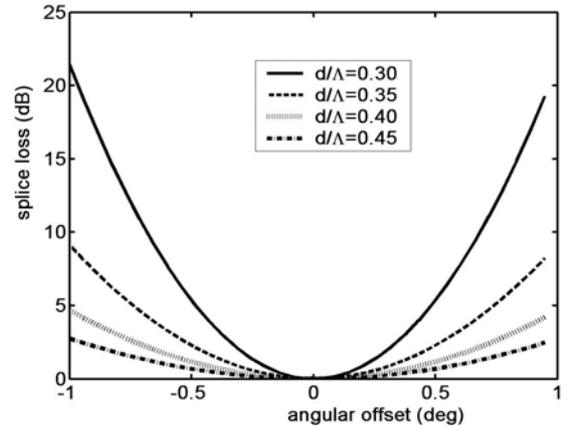

Fig.8: Splice loss in (dB) between two identical PCF versus angular offset for the parameters as in Fig. 3.

The splice loss due to angular offset for increasing ratios d/Λ is shown in Fig. 8. As it is seen, PCFs become more sensitive as the hole diameter increases or the pitch Λ decreases.

At 1.55 μm, an angular offset of 1°, results in 2.5 dB penalty for PCFs with d/Λ=0.45 and 10 dB penalty for a PCF with d/Λ = 0.35. This sensitivity of PCFs to angular misalignment can be explained by the reduction of numerical aperture or equivalently the increase of directivity of the outgoing beam from the end-face of the PCF, as the ratio d/Λ increases. Therefore, during the design and manufacturing process of optoelectronic components using PCF-technique, the angular offset issue must be handled properly.

In this subsection, we present results on splice losses between a SMF and a PCF. For SMF to PCF the transverse offset may be affected during splicing owing to the refractive index mismatch and differing surface tensions of the two fibers.

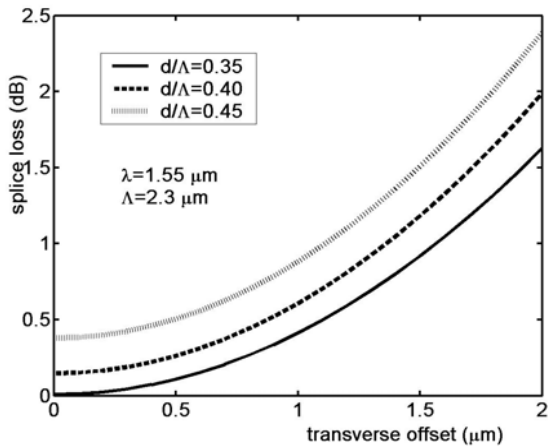

Fig.9: Splice loss in (dB) between a SI-SMF with diameter 3 μm, Δ=0.01% and a PCF with Λ=2.3 μm versus transverse offset for different hole diameters d.

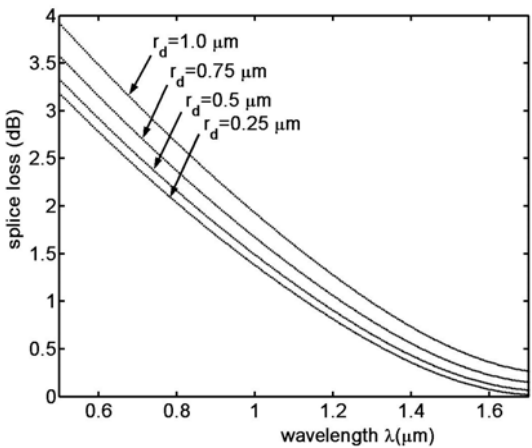

Fig. 10: Wavelength dependence of splice loss between a SI-SMF with diameter 3 μm, Δ=0.01% and a PCF with Λ=2.3 μm versus transverse offset for different offsets $r_d$.

In Fig.9 we plot the splice loss versus transverse offset for splicing a conventional SI-SMF (α = 1.5 μm, Δ=0.01 %) to a PCF. We see that the splice loss increases with hole diameter. Assuming negligible transverse offset, the optimum loss due to mode field radius mismatch between the SMF and PCF with d/Λ=0.45, is 0.378 dB.

Finally, we have investigated the wavelength dependence of splice loss due to transverse offset for splicing SI-SMF to PCF. As it is seen in Fig. 10, the loss decreases when the wavelength increases and is less than 0.5 dB, when the wavelength is in the range 1.3 μm < λ < 1.7 μm.

## 6 Conclusion

In this paper, we present analytical calculations of macrobending and splice losses for index-guiding PCFs. The calculations were based on the effective index model (EIM) and on the Gaussian beam propagation theory.

To deduce macrobending loss in PCFs, we apply the standard radiation model for bend conventional fibers, making a full transformation of conventional SIF parameters such as Δ, W, and V to fiber parameters appropriate to index-guiding PCFs. It is shown that macrobending loss can be controlled by varying the fiber parameters. The spectral window, in which a PCF can operate while remaining single mode, is observed for different values of relative air-hole sizes. The effect of tailoring the size of air-holes and pitch on spectral window is also investigated and it is found that the range of operating wavelengths for minimum bending loss, increases as the hole size increases. PCFs seems to be more bend resistant than the conventional fibers, for a particular value of bend radius and operating wavelength range.

The Gaussian beam propagation theory, for the fundamental modal field, leads to simple closed-form expressions for the splice losses produced by transverse, longitudinal and angular offsets. The effective spot size for different configurations of PCFs is obtained, which is used for computing the splice losses. The wavelength dependence of the cladding index of the PCF has been taken properly into account.

The splice losses due to transverse, longitudinal as well as angular offsets are strongly dependent on the geometric characteristics of the PCF. We found a high sensitivity of PCF-splicing to angular offset and thus angular offsets must be handled properly in designing PCF-components.

Loss due to splicing PCFs to standard single-mode fibers has also been investigated because of its importance to broad band applications. It is our aim to extend the calculations when the three types of misalignment (transverse, longitudinal and angular) are simultaneously present. We expect that the above study on macrobending and splice losses maybe helpful in the design and development of tele-communication and sensor systems based on PCFs.